# On the Size of the Mission Suite Enabled by NASA's Deep Space Network

## T. Joseph W. Lazio*

*Jet Propulsion Laboratory, California Institute of Technology, United States, Joseph.Lazio@jpl.nasa.gov*
\* Corresponding Author

## Abstract

The Deep Space Network (DSN) is the primary means of commanding, tracking, and receiving data from all of NASA's deep space missions, as well as a number of deep space missions operated by other international space agencies. The current number of missions enabled by the DSN is approximately 40 missions, but there has been concern about the level of "over-subscription" of the DSN, namely that the number of missions currently using the DSN is larger than can be enabled reasonably. This manuscript assesses the maximum number of missions that could be enabled, based on recent performance and with the constraint that the total number of hours used per week does not exceed the available number of DSN antenna-hours. Three different models are considered, and the maximal number of missions that could be enabled ranges between approximately 40 missions and 70 missions, assuming that there continues to be approximately six Mars missions and that those Mars missions continue to make use of the DSN's multiple spacecraft per antenna (MSPA) capability. Crucially, the conclusion that an approximately 50% growth in the DSN mission suite rests on the assumption that the DSN antennas are "interchangeable," but they are not, with some spacecraft able to use only certain antennas. Efforts to make the DSN antennas more "interchangeable," primarily in their transmitter and receiver suites, would be an effective means of ensuring expanded capability. Additional findings from this work are that, while additional use of the MSPA capability might appear to be a promising means for increasing the mission suite, there appear to be no locations in the Solar System, other than Mars, for which it would be effective.

## 1. Introduction

There is an increasing level of concern about the extent to which NASA's Deep Space Network (DSN) is "over-subscribed" and unable to meet even the existing needs for NASA's deep space fleet [e.g., *Origins, Worlds, and Life: A Decadal Strategy for Planetary Science and Astrobiology 2023–2032*, 1]. Figure 1 provides one illustration of how the number of missions being enabled by the DSN has increased over the past 40 years. As one measure of demand, the number of DSN antennas has remained essentially constant for the past 25 years while the number of spacecraft tracked has increased by more than 50%.

If the level of over-subscription is too high, then even NASA's existing deep space fleet is returning too little data, given the investment in those missions, and there is the concomitant risk that ambitious future science missions will be affected. At minimum, a too-high over-subscription rate could mean that fewer data are returned from future missions, in the worst case, scientists and mission planners design spacecraft and missions with lower than ideal science data returns.

By the same token, some level of over-subscription is desirable. As Figure 2 illustrates, the distribution of spacecraft in the sky changes over time so that sometimes they are distributed approximately uniformly and sometimes they are more "clumped." If the number of DSN antennas equalled the number of spacecraft, during times when the spacecraft are "clumped" in the sky, there inevitably would be some DSN antennas that would be idle.

This concept of optimizing demand versus resource availability is common to many infrastructures, with notable examples being roadways and computer networks. For example, the same roadway that is heavily congested, with low traffic speeds, at 5:00 p.m. ("rush hour") is nearly empty at 2:00 a.m. (Figure 3). If the roadway has sufficient capacity to handle "rush hour" traffic, then it has far too much capacity to handle early morning traffic. There is a rich literature that considers various metrics and figures of merit for optimization, potential incentives, and trade-offs between infrastructure capability and other considerations.

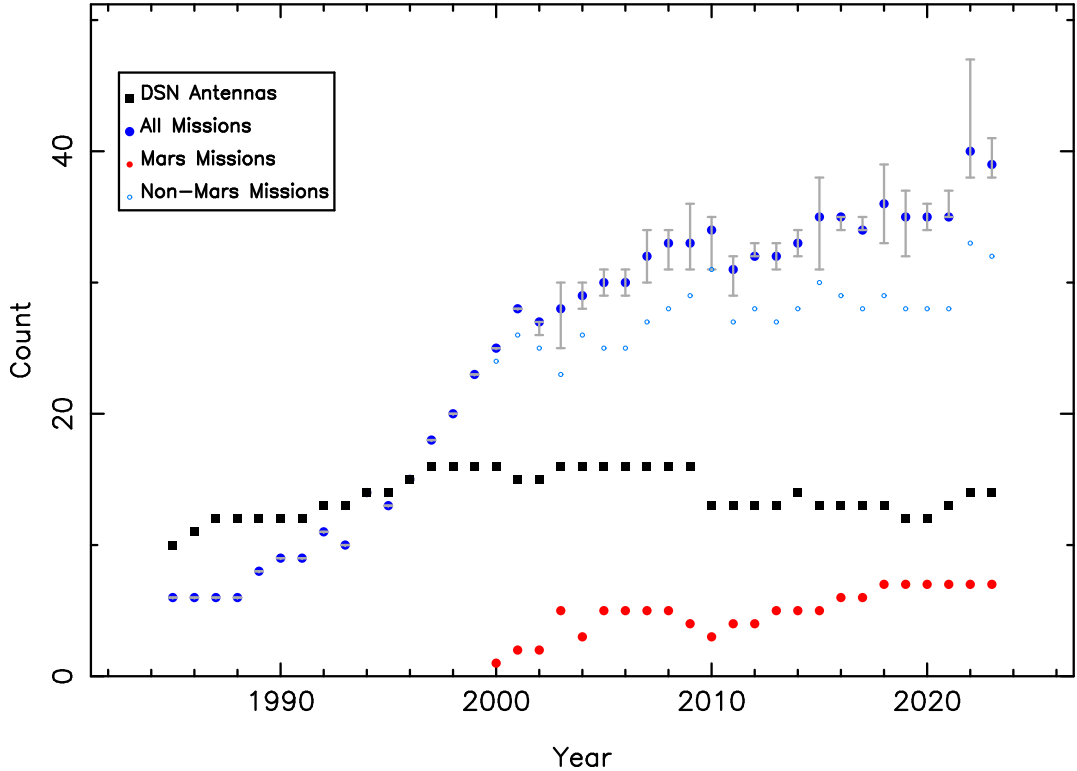

Figure 1. The number of deep space missions being enabled by NASA's Deep Space Network (DSN) has increased over the past decades, leading to concerns that the network is "over-subscribed." Shown are the number of DSN antennas in service (black squares) and the average number of spacecraft being tracked (blue circles). If the number of spacecraft varied during a year, a gray bar indicates the range in the number of spacecraft tracked for that year. Anticipating subsequent discussion, starting in 2000, missions at Mars are indicated by red circles and non-Mars missions are indicated by (small open) light blue circles.

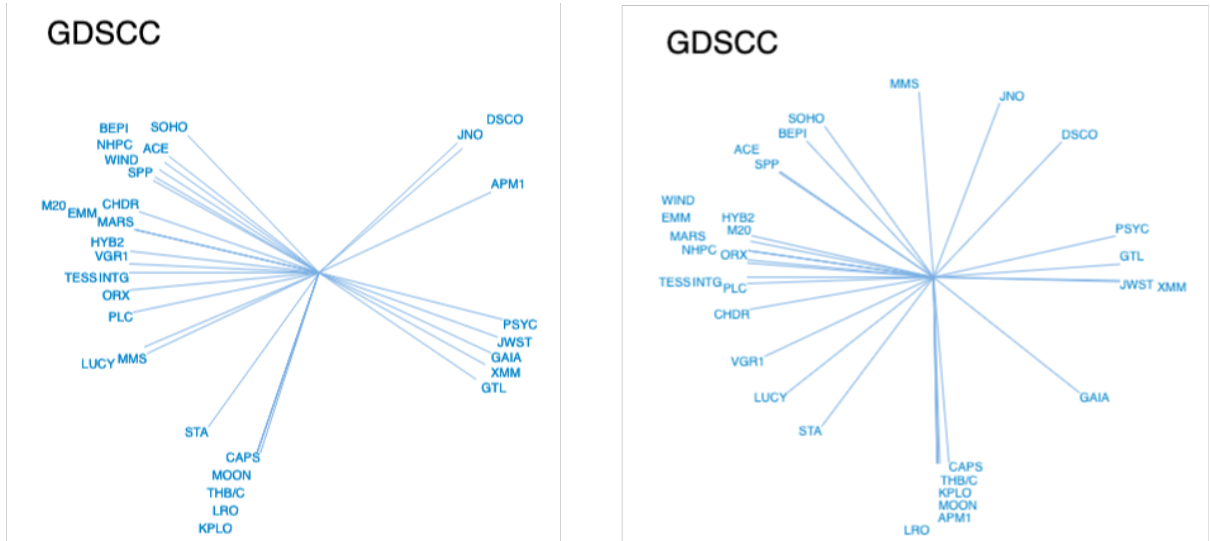

Figure 2. The distribution of spacecraft in the sky changes over time, leading to an inevitable over-subscription of DSN antennas during a typical year. Both panels show the distribution of the various spacecraft in azimuth as seen from the Goldstone Complex during 2024, with each radial bar representing one spacecraft. (*left*) Week 1, an illustration of a week in which spacecraft are "clumped" in limited ranges of azimuth and there is likely to be over-subscription; (*right*) Week 5, an illustration of a week in which spacecraft are distributed more evenly in azimuth with considerably less over-subscription.

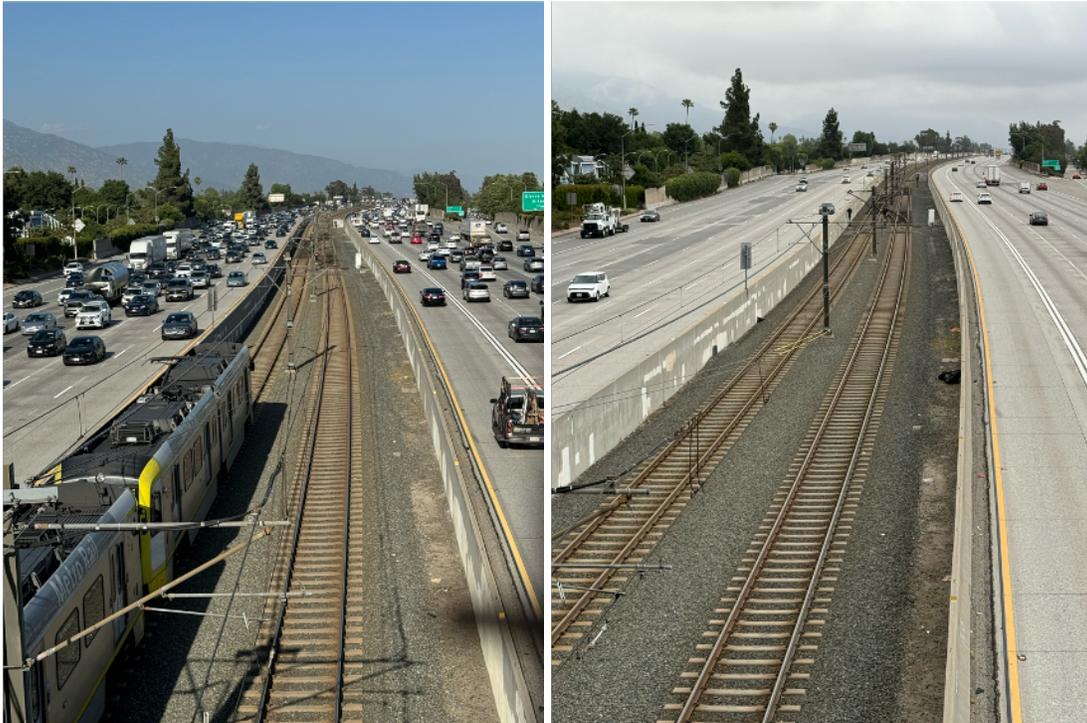

Figure 3. Over-subscribed? (left) An over-subscribed road, with vehicles moving well below the speed at which the road is designed to handle. This picture was taken on a Thursday at approximately 5:30 p.m. (right) An under-subscribed road, with considerable distance between vehicles and vehicles moving at or above the stated speed limit. This picture was taken on a Sunday morning at approximately 7:30 a.m. As is likely obvious, the "over-subscribed" and "under-subscribed" roads are the same road.

## 2. The Increasing DSN Load

Returning to Figure 1, there are three, non-exclusive explanations for the fact that the number of missions has been increasing over time while the number of antennas has remained essentially constant.

**Decreased Antenna Time per Mission**  Given that the number of DSN antennas has remained essentially constant, the total number of antenna-hours has remained essentially constant as well. One means by which more science missions can be enabled would be if the amount of DSN time that each mission receives has been decreasing with time.

**Increased DSN Efficiency**  Given that the number of antenna-hours has remained essentially constant, if those antennas-hours can be used more efficiently, then more time spent tracking science missions should be possible. As one simple example, there are required minimum durations for configuring an antenna to track a spacecraft, depending upon the details of the track. If those minimum durations can be reduced, then there should be more time available to track science missions. A prime example of DSN efficiency has been the introduction of the multiple spacecraft per aperture (MSPA) technique, which allows science data from up to four spacecraft visible within a single antenna's beam to be downloaded simultaneously.

**Decreased DSN Maintenance**  Like all mechanical devices, the DSN antennas and associated infrastructure require regular maintenance to remain working. One approach to enable more science missions would be to reduce the amount of time spent on maintaining DSN antennas.

There is some ambiguity in how to distinguish the second two categories. For instance, if a maintenance task can be conducted more efficiently, then it might be able to be conducted less frequently, allowing some of that time previously devoted to that task to be used instead for tracking science missions. Another possibility might be if more robust components or supplies begin to be used, some maintenance tasks might be able to be conducted less frequently, with no effect on antenna performance.

I now consider the extent to which each of these three possibilities might be operative and what is known about current performance.

*2.1 Decreased Antenna Time per Mission*

Table 1 shows the average numbers of hours per week for a number of deep space missions enabled by the DSN, across the three Divisions within the Science Mission Directorate that execute such missions.

The number of missions considered in Table 1 is significantly smaller than the total number enabled currently by the DSN or shown in Figure 1. There are two reasons for this reduced set of missions. First, I do not show the Mars missions. Because of the Mars Relay Network, time tracking orbiters at Mars can be used for downlinking science data from both the Mars orbiter itself and a Mars surface asset (rover or lander). Moreover, as noted above, the DSN's MSPA technique means that multiple Mars mission can share the same DSN antenna. It was beyond the scope of this document to attempt to separate all of the tracks at Mars into the respective missions.

*Table 1. Average Weekly Duration for DSN Tracking of Deep Space Missions*

| Mission | FY18 | FY22 | FY23 |
|---|---|---|---|
| | (hr) | | |
| **Astrophysics Missions** | | | |
| *Chandra* | 44.8 | 41.2 | 42.0 |
| TESS | 13.9 | 11.2 | 11.6 |
| **Heliophysics Missions** | | | |
| Advanced Composition Explorer | 37.2 | 36.1 | 34.9 |
| Magnetospheric Multiscale (MMS) | 84.4 | 72.7 | 66.4 |
| STEREO A | 63.5 | 49.2 | 44.8 |
| Voyager Interstellar Mission | 125.1 | 132.2 | 147.8 |
| Wind | 31.2 | 32.6 | 31.7 |
| **(*Non-Mars*) Planetary Science Missions** | | | |
| Juno | 72.8 | 74.4 | 85.0 |
| New Horizons | 41.3 | 42.9 | 37.7 |
| Lunar Reconnaissance Orbiter | 25.0 | 18.5 | 19.5 |
| Origins Spectral Interpretation Resource (OSIRIS-REx) | 51.1 | 36.8 | 75.8 |
| Tracking hours are reported only for STEREO A because, while there were a small number of hours spent tracking STEREO B, these were residual recovery efforts. | | | |

Second, over time scales longer than approximately five years, the actual missions enabled by the DSN can change significantly. For instance, the *Cassini* mission ended just before the beginning of (fiscal year) 2018, while the Parker Solar Probe began during the latter half of (fiscal year) 2018.

From Table 1, there is no clear evidence that more science missions are being enabled by a trend of reduced time per mission. The number of missions has grown from 35 in (fiscal year) 2015 to 40 in (fiscal year) 2023, an increase of approximately 15. For some missions, there is a trend of reduced DSN time over the last five years, such as TESS and MMS. Many of the other missions have been nearly constant or only slight reductions, and there are some missions for which the number of DSN hours have increased.

*2.2 Increased DSN Efficiency*

There is some evidence to indicate that the DSN has introduced efficiencies that could result in an increased number of hours for science missions. In addition to the MSPA technique noted above, DSN operators now handle multiple missions, whereas they now handle up to three (formally "multiple links per operator").

Another more recent example is that small reductions ($\approx 10$ min.) in the amount of time that it takes to configure a DSN antenna for various kinds of operations have been able to be implemented. While the individual changes are small, the projected savings is approximately 60 antenna-hours per week.

*2.3 Decreased DSN Maintenance*

At the time of writing, there are insufficient data available to assess the extent to which reduced maintenance might be part of the explanation for the increased number of missions enabled. However, as antennas and related infrastructure age, a reasonable expectation is that the amount of time required for maintenance should increase.

## 3. Projections for the Future

This section considers what might be the maximum number of missions that could be enabled by the DSN, without significant reductions to the amount of time per mission. I consider three different scenarios, designed to illustrate the potential range of answers.

However, it is likely that there is no single unique answer. Beyond the simple fact that the suite of missions enabled by the DSN has changed over time, and can be expected to change over time, in response to scientific discoveries, there is the more important factor that, because DSN time is a shared resource, any degradation will be a slow process. That is, there is no critical threshold or "cliff." I return to this point below (§3.1).

As an initial estimate, this analysis makes the following assumptions

- The DSN has 14 antennas, of which 12 antennas can be used for tracking deep space missions other than those at Mars. Experience has shown that the effective number of antenna-hours devoted to missions at Mars amounts to between two and three antennas. Indeed, assuming only effectively two antennas tracking Mars may be optimistic, as more recent estimates find this value to be closer to three antennas.

- Each antenna is operational for approximately 75% of the time (126 hr per week), comparable to recent experience.

- No distinction is made between the 34 m and 70 m antennas. While one might think that the larger diameter of the 70 m antennas would result in higher signal-to-noise ratios, thereby yielding higher data rates, in turn yielding less time required to obtain a given data volume, there are other considerations. For instance, the 70 m antennas may be required for more distant spacecraft (Voyager 1 and 2 or New Horizons) in order to compensate for the weakness of their signals, thereby avoiding the use of an array of antennas, which would affect many more missions.

- There is no use of the DSN MSPA technique, other than for Mars missions. Given both the current mission suite and the likely future mission suite, other than for Mars, there are no other directions for which the region of interest is sufficiently small that the MSPA technique can be used on a routine basis. Section 4.1 discusses the use of the MSPA technique further.

Accordingly, the total number of DSN antenna hours available per week is

$$T_{tot} = N_{ant} \times 168\eta \text{ hr}$$
$$= 1512 \text{ hr} \left(\frac{\eta}{75\%}\right)\left(\frac{N_{ant}}{12}\right) \hspace{3cm} 1)$$

with $\eta$ being the (fractional) operational availability of the time that a DSN antenna can be used and $N_{ant}$ being the number of DSN antennas used for *non-Mars* missions. The second line of eqn. (1) uses the values described in the assumptions above. An obvious potential future investigation might be to consider the differences between the 34 m and 70 m sub-networks, for instance, if effective operational availability $\eta$ between the two sub-nets differed significantly.

The total number of required hours for (non-Mars) science missions is then

$$T_{science} = N_A T_A + N_H T_H + N_{PS} T_{PS}, \hspace{3cm} (2)$$

where the subscript A stands for Astrophysics missions, H stands for Heliophysics missions, and PS stands for the *non-Mars* Planetary Science missions. Clearly, this set of missions could be expanded, for instance, if a significant number of missions from the Division of Biological & Physical Sciences began to use the DSN. However, this set capture the main users, to date, of the DSN. Finally, and somewhat obviously, there is the requirement that the total number of hours used must not exceed the total number available, $T_{science} \leq T_{tot} = 1512$ hr.

For the scenarios considered, Table 2 summarizes average and median weekly amounts of DSN time for missions from these Divisions within the Science Mission Directorate along with the adopted values. Clearly, the sample size is relatively limited, but it is notable that, based on past performance, an assumption of 8 hr of DSN time per days for most days of a week is reasonable.

*Table 2. Notional Weekly Duration for DSN Tracking of Deep Space Missions*

| | **FY23** | | **Adopted Value** |
|---|---|---|---|
| | **Average** | **Median** | |
| | **(hr)** | | |
| Astrophysics $T_A$ | 26.8 | 26.8 | 30.0 |
| Heliophysics $T_H$ | 61.5 | 44.8 | 48.0 |
| (*non-Mars*) Planetary Science $T_{PS}$ | 60.1 | 75.8 | 64.0 |

### 3.1 Scenario 1: Equal Time Per Mission

For the first, somewhat simplistic analysis, I consider a case in which $T_A = T_H = T_{PS} = 48$ hr. That is, all missions have an equal usage of DSN time, with a value that is in the approximate mid-range of those illustrated in Table 2.

In this case, it is straightforward to reach the conclusion that $N_A + N_H + N_{PS} < 31$. Strikingly, if this illustrative upper limit is combined with a characteristic number of Mars missions, the total number is 36 missions ($\approx 31$ non-Mars missions + 7 Mars missions), the resulting number is only slightly lower than the DSN's current mission suite ($\approx 40$ missions, Figure 1).

Returning to the discussion at the beginning of this section, an illustration of the gradual degradation to science missions, consider the scenario in which the DSN enables 31 missions, each receiving 48 hr per week. Adding another mission, bringing the mission suite to 32 missions, could be accommodated by reducing the time per mission to 46.5 hr per week, a reduction of 3.1% per mission. Is this level of reduction acceptable? Even if an individual mission might not find such a reduction acceptable, NASA's Science Mission Directorate and the science community might accept such a reduction based on the science return of the mission to be added. Adding another mission, bringing the mission suite to 33 missions, would reduce the time per mission to 45.1 hr per week, equivalent to 6% per mission. Again, is this level of reduction acceptable?

### 3.2 Scenario 2: Mission-Specific Times

For this scenario, I adopt the notional values from Table 2. Ignoring "edge cases," such as setting one of the parameters $N_A$, $N_H$, or $N_{PS}$ to 0, it is still clear that there is no single, unique answer.

Historically, few Astrophysics missions have used the DSN, in part because effective missions could be executed below geosynchronous orbit (GEO), for which the DSN is not a good solution. While more Astrophysics missions may be placed into orbits around Sun-Earth Lagrange point 2, akin to *JWST*, the number of Astrophysics missions using the DSN likely will remain smaller than those from the Heliophysics or (non-Mars) Planetary Science Divisions.

Figure 4 shows the resulting ranges of Heliophysics and (non-Mars) Planetary Science missions that can be enabled for different assumptions about the assumed number of Astrophysics missions also being enabled. Illustrated are two examples, one in which the number of Astrophysics missions enabled by the DSN is comparable to the current number and one in which it approximately doubles. In all cases, the total number of (non-Mars) missions is fewer than 35 missions.

Assuming approximately seven Mars missions, the total number of missions in the DSN mission suite would be 42 missions, comparable to the current DSN mission suite.

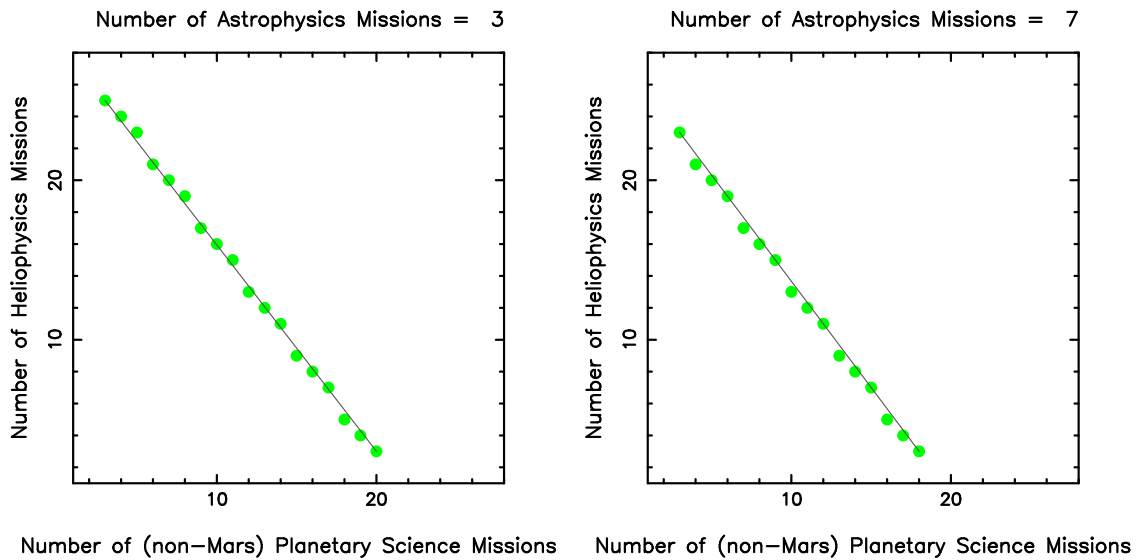

Figure 4. Projected ranges of missions from the Heliophysics and (non-Mars) Planetary Science enabled by the DSN for the assumed number of Astrophysics missions, using the notional DSN times specified for each mission time in Table 2.

### 3.3 Scenario 3: Usage-Based Mission Scheduling

In this scenario, the DSN time allotted to a mission is based in some measure on a "priority."[1] While many approaches could be adopted, I choose a simple one in order to illustrate the potential results. I assume that missions are assigned "high," "intermediate," or "low" usage in order to reflect the typical amount of time that a mission might receive per week. I stress that "low" usage is *not* equivalent to "no" priority. All operating missions require some contact with the DSN, and lack of contact could result in the end of a mission. Moreover, the geometry (Figure 2) may result in times such that only low usage missions are able to be tracked at some times of the day.

---

[1] The concept of priorities is a standard metric for NASA's Astrophysics missions—proposals for using a space telescope for a certain number of hours to conduct a set of proposed observations are judged according to their scientific merit. Results are commonly that some proposals receive the requested observing time, some proposals may receive less than the full request of observing time, and some proposals may be declined completely, receiving no observing time.

For this scenario, the total number of hours for (non-Mars) science missions is

$$T_{science} = N_{hi}T_{hi} + N_{mid}T_{mid} + N_{lo}T_{lo}, \tag{3}$$

where $N_{hi}$, $N_{mid}$, and $N_{lo}$ are the number of high-, intermediate-, and low-usage missions, respectively, and $T_{hi}$, $T_{mid}$, and $T_{lo}$ are the average number of hours for those missions, respectively. Somewhat by definition, $T_{hi} > T_{mid} > T_{lo}$, and $N_{hi} < N_{mid} < N_{lo}$.

Figure 5 summarizes the usage of the DSN by various missions during the (calendar) year 2023.[2] Guided by Figure 5, I take $T_{hi} = 65$ hr, equivalent to a mission making use of one-half of the (effective) number of hours on a DSN antenna in a week, and $T_{mid} = 30$ hr, equivalent to a mission making use of one-quarter of the (effective) number of hours on a DSN antenna in a week. Finally, I choose $T_{lo} = 5$ hr. In practice, this time may be spread across antennas as a mission typically is tracked by antennas at different DSN Complexes.

For the purposes of illustration, I consider $N_{hi} = 5$. It is straightforward to determine that a maximum number of missions would be obtained with 34 intermediate-usage and 33 low-usage missions, for a total DSN science mission suite $N_{science} = 72$ using a total DSN time per week $T_{science} = 1510$ hr. Different choices for the various parameters likely would lead to different results, but this simple exercise does illustrate that there are scenarios in which the number of missions enabled by the DSN could increase by potentially as much as 50%.

---

[2] Consistent with historical precedent and international science engagement, the DSN routinely tracks missions from international space agencies, such as the European Space Agency (ESA) or the Japanese Aerospace Exploration Agency (JAXA), a practice that occurred in 2023.

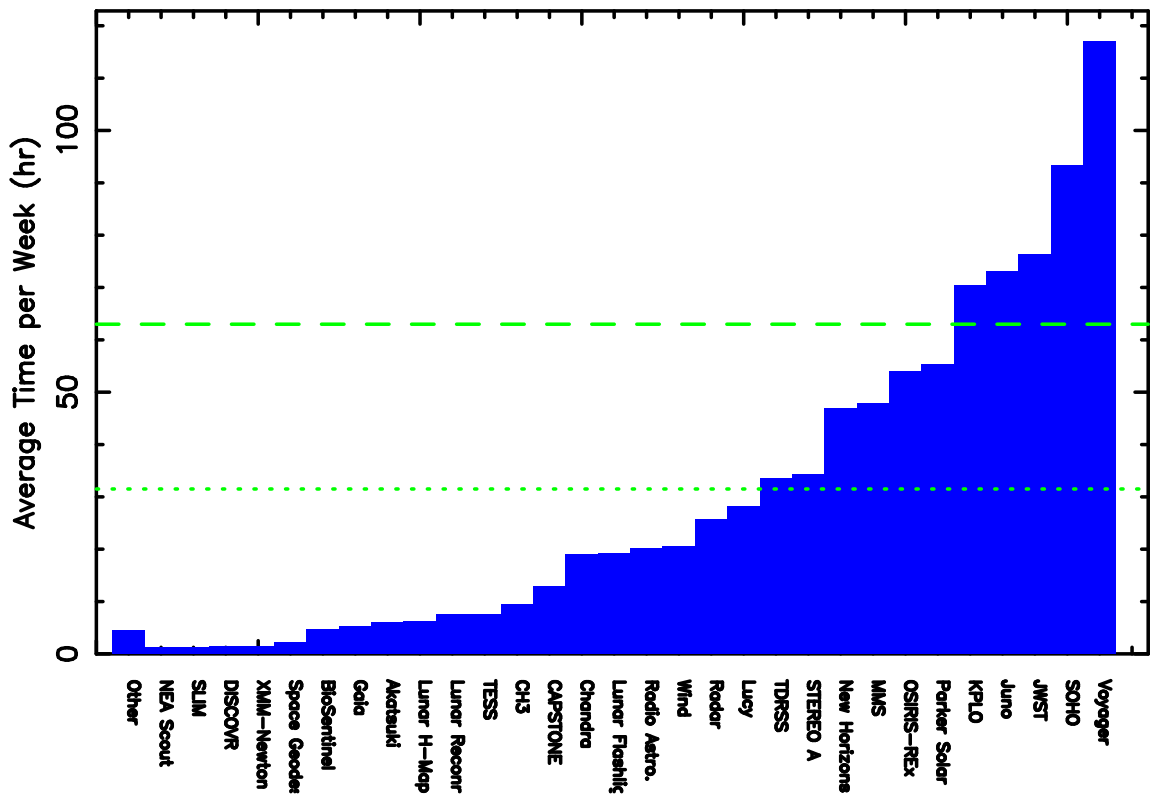

Figure 5. Average time (in hours per week) used by the missions in the DSN mission suite during calendar year 2023. For missions that include more than one spacecraft, the average time shown accounts for DSN tracks on all spacecraft in that mission. The category "Other" (right-most mission shown) includes all missions for which the average time per week is less than 1 hr. The (horizontal) dashed line marks 84 hr per week, equivalent to one-half of the total hours in a week, and the (horizontal) dotted line marks 42 hr per week, equivalent to one-quarter of the hours in a week. These values are used in the analysis for "high" and "intermediate" mission usage of the DSN. Not shown are the tracks for Mars missions or time for maintenance, and some of the missions listed are from international partners. ("Radar" = DSN radar observations; "Radio Astro." = DSN Radio Astronomy observations)

## 4. Additional Topics

In this section, I expand upon various topics noted earlier.

### 4.1 Multiple Spacecraft per Antenna (MSPA)

As noted above, the MSPA technique has introduced considerable efficiencies for Mars missions, enabling science data from up to four spacecraft at or on Mars to be downloaded simultaneously. However, the potential for the MSPA technique to be used for other missions is limited, as there are no other directions for which the region of interest is sufficiently small that the MSPA technique can be used. For reference, the antenna power pattern for a 34 m antenna receiving science data at 8.45 GHz (X band) is 0.066°; the power pattern for higher frequencies (K- or Ka bands) is smaller. The two regions noted most commonly as relevant for the use of the MSPA technique are the Moon and Sun-Earth Lagrange points. However, in both cases, the regions of interest are (much) larger than a DSN antenna power pattern.

The Moon itself is 0.5° in diameter, and so-called near-rectilinear halo orbits (NHROs) can take spacecraft up to 10° from the Moon. Tellingly, there currently are four spacecraft orbiting the Moon—Lunar Reconnaissance Orbiter (LRO); two Acceleration, Reconnection, Turbulence and Electrodynamics of the Moon's Interaction with the Sun (ARTEMIS) spacecraft, ARTEMIS-P1 and ARTEMIS-P2 (known previously as Time History of Events and Macroscale Interactions during Substorms [THEMIS] B and THEMIS C); and the Cislunar Autonomous Positioning System Technology Operations and Navigation Experiment (CAPSTONE). A review of the DSN schedule over the past few years found no instance in which MSPA was used for any combination of these four spacecraft. However, there were instances in which ARTEMIS-P1 (THEMIS B) and ARTEMIS-P2 (THEMIS C) were tracked sequentially on the same antenna at a Complex, and there were simultaneous passes on ARTEMIS-P2 (THEMIS C) and CAPSTONE using different antennas at the same Complex.

The other notable locations for spacecraft are the Sun-Earth Lagrange 1 (S-E L1) and Sun-Earth Lagrange 2 points. Both Lagrange points are located along the Sun-Earth line, approximately $150 \times 10^6$ km from the Earth. The S-E L1 point is located between the Sun and Earth and is a favorable location for Heliophysics missions to observe the Sun while the S-E L2 point is more distant from the Sun than the Earth and is a favorable location for Astrophysics missions to observe the Universe.

At the time of writing, there are three spacecraft at the S-E L1 point, the Advanced Composition Explorer (ACE), the Solar and Heliospheric Observatory (SOHO), and the Deep Space Climate Observatory (DSCOVR). Until the recent conclusion of the operations of the Gaia spacecraft, there were two spacecraft at the S-E L2 point, the other spacecraft being the *James Webb Space Telescope* (*JWST*). After the launch of the Nancy Grace Roman Space Telescope (notionally by 2027 May), there will once again be two spacecraft at the S-E L2 point.

Orbits at either S-E L1 or S-E L2 are not stable. Rather than being "at" a Sun-Earth Lagrange point, spacecraft are in "halo" orbits around the appropriate Lagrange point. For instance, *JWST*'s offset from S-E L2 varies from $0.25 \times 10^6$ km to $0.83 \times 10^6$ km, corresponding to offset angles of up to 30° from the S-E L2 point itself.

There are few other destinations within the Solar System where there are multiple spacecraft currently or where multiple missions are planned. The notable exception might be Venus, for which there are currently three planned missions, the Venus Emissivity, Radio Science, InSAR, Topography, and Spectroscopy (VERITAS) mission, the Deep Atmosphere Venus Investigation of Noble gases, Chemistry, and Imaging (DAVINCI) mission, and the EnVision mission. Assuming that all three missions maintain their current schedules, there could be two or three spacecraft at Venus by the middle of the 2030s decade.

*4.2 Receiver Suites*

An underlying assumption for all of the analyses conducted here is that the DSN antennas are identical. While there is the obvious difference between the 34 m-diameter and 70 m-diameter antennas, even the 34 m-diameter antennas are not identical. This difference is most apparent in considering the receiver configurations for the various 34 m-diameter antennas [2]. The DSN antennas have capabilities in four frequency bands, identified as "Space Research Services" by the International Telecommunications Union (ITU). These four bands are identified by (historical) letter codes as S band ($\approx 2.2$ GHz), "X band" ($\approx 8$ GHz), K band ($\approx 26$ GHz), and Ka band ($\approx 33$ GHz). The frequencies used for transmitting commands and receiving data are slightly different, and there are further distinctions depending upon whether the spacecraft is within 2 million km of the Earth, but these are of secondary importance.

The DSN antennas have a common, or nearly common, implementation of X band capabilities, such that antennas effectively are identical and interchangeable. That is not the case at other frequencies, with some antennas have S- and K band capabilities, but not Ka band capabilities, or vice versa. Even within a specific frequency band, capabilities can differ, most notably with the transmitter powers being different for different antennas. The fact that the antennas are not identical and effectively interchangeable reduces the possibility to optimize mission usage, as some missions can use only a limited subset of antennas with other antennas being able to use only a different limited subset.

Not only do these differences limit the ability to optimize across the full SMD mission suite, there is anecdotal evidence that these differences can affect mission development. The most common experience is that a mission concept will be formulated assuming that any 34 m-diameter DSN antenna can be used, only to discover during later development that this assumption is invalid, requiring rework or modifications to the spacecraft telecommunications subsystem, mission operations, or both.

*4.3 Other Expansions in Capacity*

A long-standing goal has been to have spacecraft use K- and Ka-bands, rather than X band, for transmitting science data [e.g., 3,4]. The motivation for this change is that there is the potential for a factor of four improvement in performance, specifically in terms of the transmitted data rate. This potential increase in data rate could have one of two consequences for the DSN mission suite—if downlinked data volumes remain unchanged, the amount of DSN time required per mission would decrease, or downlinked data volumes could quadruple, which would not change the amount of DSN time per mission.

There is some evidence to suggest that the latter possibility would be the more likely. First, missions already limit, sometimes dramatically, the amount of science data downlinked, potentially precluding future scientific discoveries or advances. Specific examples of such limited science data transmission include the *Kepler* mission, for which, due to "to data storage and transmission limitations, only about 6% of the 96 million pixels [acquired were] stored for eventual transmission to the ground" [5], and the High Resolution Imaging Science Experiment (HiRISE) on the Mars

Reconnaissance Orbiter (MRO), which has covered "just a few percent of" Mars even after 16 years of operation [6]. Second, innovative and efficient uses of science instruments on a spacecraft can be developed, such as enabling scientifically-valuable observations to be acquired simultaneously by a primary and secondary detector. The most recent such example is the development of the "pure-parallel" observing mode for the *JWST* [7,8].

Finally, the *JWST* uses K band and Parker Solar Probe uses Ka band for science data transmission. These two missions are among those obtaining the largest allocations of DSN time (Figure 5).

## 6. Conclusions

Motivated by continuing concerns that NASA's Deep Space Network (DSN) is "oversubscribed," I have considered the maximum number of science missions that the DSN can enable, based on recent performance and given the current set of antennas. This analysis has considered the question from the perspective of number of DSN antenna-hours available in a typical week, and I have considered three different scenarios—one in which all missions receive the same amount of DSN antenna time, one in which the set of missions somewhat reflects historical use by the different Divisions within NASA's Science Mission Directorate (SMD), and one in which some ("high usage") missions receive more time than other ("intermediate usage" and "low usage") missions. This set of scenarios is not intended to be comprehensive, but it should illustrate the potential range for the size of the mission suite enabled by the DSN.

This analysis has not included any of the missions to Mars, because there are so many of them currently and the Mars Relay Network makes determining a specific number of DSN antenna-hours per mission difficult to determine. Should the number of Mars missions decrease in the future such that Mars is not a "special" location in the Solar System, this analysis would be worth repeating.

I find that the number of total missions that the DSN can enable ranges between approximately 40 missions and 70 missions. This finding assumes that the number of Mars missions remains similar to recent experience, namely approximately six Mars missions. My conclusion is that current mission suite, of approximately 40 missions, is *not* maximal, as further efforts to enable the DSN mission suite potentially to grow by as much as 50%. However, this analysis also has made an implicit assumption that all 34 m DSN antennas are identical, which they are not. A straightforward approach to ensuring that the DSN can continue to enable a full mission suite would be to ensure that all 34 m antennas have identical capabilities to the extent possible.

This approach necessarily leads to a certain amount of over-subscription as not all missions will be able to obtain the notional amounts of DSN time when a large number of spacecraft are "clumped" in a relatively limited portion of the sky. Alternate approaches could be considered, such as requiring that the minimal amount of DSN time obtained by any mission should never drop below a specific threshold or requiring that the number of "idle" DSN antenna-hours never drop below a threshold.

Unfortunately, while the Mars missions can make effective use of the DSN's MSPA capability, there are few, if any, other locations in the Solar System, for which its use likely will be as effective.

## Acknowledgements

I thank S. Lichten, B. Arroyo, S. Asmar, R. Castaño, and M. Johnston for illuminating discussions and comments that helped improve the presentation of this material. This research has made use of NASA's Astrophysics Data System. This research was carried out at the Jet Propulsion Laboratory, California Institute of Technology, under a contract with the National Aeronautics and Space Administration. Some of this material is pre-decisional information and for planning and discussion only.